
\physrev

\def\SCIPP{\centerline {\it Santa Cruz Institute for Particle Physics}
  \centerline{\it University of California, Santa Cruz, CA 95064}}
\def\ifmath#1{\relax\ifmmode #1\else $#1$\fi}
\def\half{\ifmath{{\textstyle{1 \over 2}}}}
\def\third{\ifmath{{\textstyle{1 \over 3}}}}
\def\calo{{\cal O}}
\def\call{{\cal L}}
\def\ie{{\it i.e.}}

\def\bold#1{\setbox0=\hbox{$#1$}%
     \kern-.025em\copy0\kern-\wd0
     \kern.05em\copy0\kern-\wd0
     \kern-.025em\raise.0433em\box0 }


\Pubnum={SCIPP-92/30}
\date={July, 1992}
\pubtype{}
\titlepage
\vskip 4cm

\title{\bf Spontaneous CP violation in Supersymmetric theories}
\author{Alex Pomarol}
\vskip .1in
\SCIPP
\vskip .2in

\vfill
\centerline{\bf Abstract}
We study the minimal version of the
supersymmetric standard
model with spontaneous CP breaking.
In this model, the KM matrix is real and
 contributions to $\varepsilon$ arise from
box diagrams involving squarks. We analyze the
region of the parameter space which corresponds to values for
$\varepsilon$, $\varepsilon^\prime/\varepsilon$ and the neutron electric
dipole moment (NEDM) in agreement with the experimental data.
 We show that
 the CP violating phases must be of $\calo (10^{-2})$ and the NEDM  lies
near its present experimental limit.

\vskip .1in

\vfill
\endpage

\noindent{\bf 1. Introduction}

\REF\pq{R.P. Peccei and H.R. Quinn, Phys. Rev. Lett. {\bf 38} (1977)
1440; Phys. Rev. {\bf D16} (1977) 1791.}
The origin of CP nonconservation is not yet fully understood.
One major concern in recent times has been to understand
why in  the standard model (SM)
 the CP violating phase of QCD, $\bar\theta$, is so small
(\ie, the strong CP problem).
Several solutions to this puzzle have been proposed in the
literature.
The most appealing idea is the Peccei-Quinn (PQ) mechanism\refmark\pq.

\REF\nedma{J. Ellis, S. Ferrara and D.V. Nanopoulos,
Phys. Lett. {\bf B114} (1982) 231;\hfill\break
W. Buchm\"uller and D. Wyler, Phys. Lett. {\bf B121} (1983) 321.}
\REF\nedmb{J. Polchinski and M. B. Wise,
Phys. Lett. {\bf B125} (1983) 393.}
\REF\nedmc{F. del Aguila, M.B. Gavela, J.A. Grifols, and A. M\'endez,
Phys. Lett. {\bf B126} (1983) 71;\hfill\break
D.V. Nanopoulos and M. Srednicki, Phys. Lett. {\bf B128} (1983) 61;
\hfill\break
A.I. Sanda, Phys. Rev. {\bf D32} (1985) 2992.}
\REF\nedmd{E. Braaten, C.S. Li and T.C. Yuan, Phys. Rev. Lett. {\bf 64}
(1990) 1709;\hfill\break
J. Dai \etal, Phys. Lett. {\bf B237} (1990) 216;\hfill\break
M. Dine and W. Fischler, Phys. Lett. {\bf B242} (1990) 239;\hfill\break
R. Arnowitt, J.L. Lopez and D.V. Nanopoulos,
Phys. Rev. {\bf D42} (1990) 2423;\hfill\break
R. Arnowitt, M.J. Duff and K.S. Stelle,
Phys. Rev. {\bf D43} (1991) 3085.}
\REF\nedme{A. De R\'ujula, M.B. Gavela, O. P\`ene and F.J. Vegas,
Phys. Lett. {\bf B245} (1990) 640.}
\REF\nedmf{Y. Kizukuri and N. Oshimo, Phys. Rev. {\bf D45} (1992) 1806.}
\REF\toof{G. 't Hooft, 1979 Cargese Summer Institute Lectures.}
In supersymmetric theories, however, even if one can arrange for
$\bar\theta\simeq 0$, a new problem arises.
It has been known for a long time
that  the supersymmetric extension of the SM contains
a number of  new sources of CP violation whose contribution
to the neutron electric dipole
moment (NEDM) is two or three orders of magnitude larger than the
experimental limit if the phases that parametrize
the CP violation, $\varphi$, are of
$\calo (1)$\refmark{\nedma--\nedmf}.
Thus, a fine-tuning of parameters is
necessary  such that $\varphi\lsim 10^{-2}$--$\, 10^{-3}$.
  Since such CP violating phases arise from different
sectors of the supersymmetric model, this multiple fine-tuning appears
to be totally unnatural.
In fact, it violates 't Hooft's naturalness condition which states
that a parameter
is only allowed to be very small if setting it to zero increases the
symmetry of the theory\refmark\toof.

\REF\barr{S.M. Barr and A. Masiero, Phys. Rev. {\bf D38} (1988) 366.}
\REF\nelson{A. Nelson, Phys. Lett. {\bf B136} (1984) 387;
Phys. Lett. {\bf B143} (1984) 65;\hfill\break
S.M. Barr, Phys. Rev. Lett. {\bf 53} (1984) 329; Phys. Rev. {\bf D30}
(1984)  1805.}
\REF\dann{A. Dannenberg, L. Hall and L. Randall, Nucl. Phys. {\bf B271}
(1986) 574.}
One simple and very attractive solution to this problem is to require
that CP is spontaneously broken. In this case, CP invariance is imposed
on  the initial Lagrangian and it is broken by the ground state along
with the gauge symmetry. One example in which such idea is implemented
is the supersymmetric
version\refmark\barr\ of the Barr and Nelson models\refmark\nelson.
Models of this type
require the existence of exotic superheavy fermions which mix
with the standard light fermions. At low energy, such models are
indistinguishable from the Kobayashi-Maskawa (KM) model.
Another
 example is given in ref.~[\dann] where CP is spontaneously broken
 at a high energy scale inducing  complex scalar
  mass terms at low energy.
In such a model, extra color singlet and color triplet fields are
necessary.

In this paper we analyze the minimal version of the supersymmetric
standard model with spontaneous CP violation (SCPV).
In such a model, CP violation derives from the
phases of the vacuum expectation values (VEVs) of the Higgs bosons.
The purpose of this work
is to determine whether this model
can explain the CP nonconservation
observed in the $K$--$\bar K$ system while being
consistent with the present
bounds on the NEDM.

\REF\km{F. del Aguila \etal,
Phys. Lett. {\bf B129} (1983) 77;\hfill\break
J.M. Fr\`ere and M.B. Gavela, Phys. Lett. {\bf B132} (1983) 107;
\hfill\break
J.M. G\'erard, W. Grimus, A. Raychaudhuri and G. Zoupanos,
Phys. Lett. {\bf B140} (1984) 349;\hfill\break
P. Langacker and B. Sathiapalan, Phys. Lett. {\bf B144} (1984) 401.}
It has been claimed for a long time that supersymmetric models
need the KM phase in order to explain the CP
violating phenomena\refmark\km.
 This statement is based on examining
spontaneously broken N=1 supergravity theories with a flat K\" ahler
 metric (\ie, all the scalar kinetic terms are canonical). In these
theories
$$\Delta m^2_{\tilde q}\equiv
(m^2_{\tilde q_1}-m^2_{\tilde q_2})\sim
 (m^2_{q_1}-m^2_{q_2})\, ,\eqn\dif$$
 where $\tilde q_1$ and $\tilde q_2$ are the scalar-partners (squarks)
 of the $q_1$ and $q_2$ quarks respectively. Since box diagrams involving
superpartners (superbox) are suppressed by a factor
($\Delta m^2_{\tilde q}/m^2_{\tilde q})^2$  due to the so-called
super-GIM  mechanism, their contributions to
$\varepsilon$ are negligible for $\varphi\lsim 10^{-2}$.

\REF\soni{S.K. Soni and H.A. Weldon, Phys. Lett. {\bf B216} (1983) 215.}
When  more general N=1 supergravity theories are considered,
eq.~\dif\ is no longer satisfied\refmark\soni\  and superbox
diagrams can be
phenomenologically important. In fact, in
 such theories, the squark mass matrix
is completely
arbitrary, and as a result its diagonalization is independent
of the diagonalization of the quark matrix. This
implies that the unitary
matrices, $V^a$, which characterize the Higgs or gauge fermionic-partners
(higgsino or gaugino, $\tilde\chi_a$)
 interactions with quarks and squarks, \ie,
$$\call_{int} \prop V_{ij}^a \bar q_i\tilde q_j(1-\gamma_5)
\tilde\chi_a+h.c.\, ,\eqn\alexa$$
are arbitrary.

\REF\hall{L.J. Hall and L. Randall, Phys. Rev. Lett. {\bf 65}
(1990) 2939.}
\REF\ellis{J. Ellis and D.V. Nanopoulos, Phys. Lett. {\bf B110} (1982)
44.}
\REF\fcnco{M.J. Duncan and J. Trampetic, Phys. Lett. {\bf B134} (1984)
439.}
\REF\fcnct{F. Gabbiani and A. Masiero, Nucl. Phys. {\bf B322}
(1989) 235.}
\REF\deg{M. Dine, A. Kagan and S. Samuel, Phys. Lett. {\bf B243}
(1990) 250.}
In this paper we will work within the context of such
 general N=1 supergravity theories.
Our results, however, can be easily generalized to a wide  class of
effective low-energy supersymmetric models.\foot{It
 has been recently emphasized\refmark\hall\ that the idea of
supersymmetry at the weak scale should be tested without regard to the
Planck-scale origin of any specific model.}
Following ref.~[\ellis ], we will assume approximately diagonal forms
for the super-KM matrices, $V^a$, similar to the standard
KM matrix:
$$V^a\simeq\left(\matrix{1& \calo
 (\sin\theta_c)&\calo (10^{-2})\cr
 \calo (\sin\theta_c)&1&\calo (\sin\theta_c)\cr
 \calo (10^{-2})&\calo (\sin\theta_c)&1}\right)\eqn\alexb$$
where  $\theta_c$ is the Cabibbo angle.
Even with this natural assumption, the contribution of the superbox
 diagrams to flavor changing neutral current (FCNC) processes is
too large unless there is some mass degeneracy between
squarks. Bounds on $\Delta m^2_{\tilde q}/m^2_{\tilde q}$ from FCNC
processes were studied many years ago in refs.~[\ellis,\fcnco].
A recent analysis can be found in ref.~[\fcnct].
Possible origins for
such a degeneracy have been explored in ref.~[\deg ].
\bigskip
\noindent{\bf 2. The Higgs sector of the Model}

\REF\hunter{See, for example
J.F. Gunion, G.L. Kane, H.E. Haber and S. Dawson,
{\it The Higgs Hunter's Guide} (Addison-Wesley Publishing Company,
Reading, MA, 1990) for a comprehensive review and a guide
to the literature.}
\REF\cpmssm{N. Maekawa, Preprint KUNS 1124 (1992).}
\REF\alex{A. Pomarol, Preprint SCIPP-92/19 (1992).}
\REF\exp{Particle Data Group, Phys. Rev. {\bf D45} (1992) S1.}
The minimal supersymmetric extension of the standard model (MSSM)
 requires two
Higgs doublets. The VEVs of the
two neutral scalars can be chosen real without loss of
generality\refmark\hunter\
so that CP cannot be spontaneously broken.
It has been recently claimed
that SCPV can occur in the MSSM when radiative corrections to the Higgs
potential are included\refmark{\cpmssm ,\alex}.
This, however, requires\refmark\alex\  a Higgs boson lighter
 than that permitted by the
LEP Higgs search\refmark\exp.
An extension of the MSSM Higgs sector is thus required if we want to
 have SCPV in supersymmetric theories.

\REF\hg{J.F. Gunion and H.E. Haber, Nucl. Phys. {\bf B272} (1986) 1.}
\REF\sing{J. Ellis \etal, Phys. Rev. {\bf D39} (1989) 844;\hfill\break
M. Drees, Int. J. Mod. Phys. {\bf A4} (1989) 3635;\hfill\break
P. Bin\'etruy and C.A. Savoy, Phys. Lett. {\bf B277} (1992) 453.}
Let us consider a model with two Higgs doublets $H_1\equiv
(H^0_1,\, H^-_1)$  and $H_2\equiv (H^+_2,\, H^0_2)$ with
hypercharges $Y=-1$ and $Y=1$ respectively and a complex singlet $N$.
Such a Higgs sector has been extensively studied in the
 literature\refmark{\hg,\sing}\
and provides an attractive solution to the $\mu$-problem.
The most general renormalizable and gauge invariant
 superpotential for one quark generation
is given by
$$\eqalign{W&=\third\lambda_1 N^3+\lambda_2H_1H_2N+
\half\mu_NN^2+\mu H_1H_2\cr
&+h_dH_1\tilde Q\tilde D^c+h_uH_2\tilde Q\tilde U^c\, ,}\eqn\super$$
 where $\tilde Q$ is the squark doublet, $\tilde U$ and $\tilde D$ are
  the squark
 singlets, and we have fixed
  the notation such that $H_1H_2\equiv H^0_1H^0_2
-H^-_1H^+_2$. The scalar potential in the
 supersymmetric limit is given by
$$V=\coeff{1}{2}\left[\sum_a\left(\half gA^*_i\sigma^a_{ij}
A_j\right)^2+
\left(\half g^\prime Y_iA^*_iA_i\right)^2\right]+\left|
\coeff{\partial W}{\partial A_i}\right|^2\, ,\eqn\alexc$$
where $A_i$ collectively
denotes all scalar fields appearing in the theory.
After spontaneous supergravity
breaking, new terms are induced
in the low-energy Higgs potential which softly break global supersymmetry
(SUSY). These are given by\refmark\soni
$$\eqalign{V_{soft}&=m^2_1|H_1|^2+m^2_2|H_2|^2+m^2_3|N|^2
+m^2_{12}H_1H_2\cr
&+m^2_NN^2+A_NNH_1H_2+A^\prime_NN^3+h.c.}\eqn\alexd$$

\REF\romao{J.C. Rom\~ao, Phys. Lett. {\bf B173} (1986) 309.}
CP invariance implies  that all couplings and mass parameters are real.
In order to have the desired pattern of gauge symmetry breaking, we will
assume that only
 the neutral components of the Higgs bosons develop VEVs:
$$\VEV{H^0_1}=v_1\, ,\ \ \ \VEV{H^0_2}=v_2e^{i\rho}\,
 ,\ \ \ \VEV{N}=ne^{i\xi}\, .
\eqn\alexe$$
For $\rho,\xi \not= n\pi\ (n\in {\cal Z})$, CP is
broken along with the gauge symmetry.
 The phase-dependent part of the Higgs boson potential can be written as
$$\eqalign{V(\rho,\xi)&=A\cos\xi+B\cos 2\xi
+C\cos 3\xi +D\cos\rho\cr
&+E\cos(\rho-2\xi)+F\cos(\rho+\xi)\, ,}\eqn\angle$$
where the new $A,\ B,\ C,\ D,\ E$ and $F$ quantities can be
  easily related
to the original parameters. It can be shown that there exists
 a region in the
parameter space where the minimum of the potential is at
 $\rho,\xi \not= n\pi$.\foot{This is only true for the
  most general superpotential of eq.~\super\  or for
theories beyond the minimal N=1 supergravity\refmark\romao.}
{}From eq.~\angle, we see that
 when $\xi$ is small, $\rho$ must be close to $0$ or $\pi$.
This means that only one fine-tuning, $\xi\ll 1$, will be
necessary in order that all CP violating effects are small;
 we will see that this is required by the NEDM bound.
The smallness of $\xi$ does not violate `t Hooft's
   naturalness condition.

\REF\bra{G.C. Branco, Phys. Rev. Lett. {\bf 44} (1980) 504.}
In our model, the CP violating processes will always involve
neutral Higgs
boson couplings. Prior to spontaneous gauge symmetry breaking,
 the neutral Higgs interactions with fermions
are given by (following the notation of ref.~[\hg ])
$$\eqalign{\call_{int}&=-h_u H^0_2\bar u_Ru_L-h_dH^0_1\bar d_Rd_L\cr
&- g\left(H^{0*}_1\bar{\tilde H}P_L\tilde W+
H^{0*}_2\bar{\tilde W}P_L \tilde H\right)
-\sqrt{\half}
\left(H^{0*}_1\bar{\tilde H_1}-H^{0*}_2\bar{\tilde H_2}\right)P_L
\left(g\tilde W_3-g^\prime\tilde B\right)\cr
&-\lambda_2\left(H^0_1\bar{\tilde N}P_L\tilde H_2
+H^0_2\bar{\tilde N}P_L\tilde H_1+
N\bar{\tilde  H_1}P_L\tilde H_2-N\bar{\tilde H}P_L\tilde H\right)\cr
&-2\lambda_1N\bar{\tilde N}P_L\tilde N+h.c.\, ,}\eqn\inta$$
where $P_L=(1-\gamma_5)/2$,
and the relevant neutral Higgs interactions with squarks are
given by\foot{The coefficients of the
soft-breaking terms $H_i\tilde q\tilde q$
 are in principle arbitrary\refmark\soni. In
agreement with standard theoretical
prejudices\refmark\hall, we will assume
 that these coefficients are
proportional to the Yukawa coupling $h_q$. Otherwise contributions to
the NEDM will be too large.}
$$\eqalign{\call_{int}&=h_u\left(\lambda_2H^{0*}_1N^*+\mu H^{0*}_1+m_6
A_uH^0_2\right)
\tilde u^*_R\tilde u_L\cr
&+h_d\left(\lambda_2H^{0*}_2N^*+\mu H^{0*}_2+m_6
A_dH^0_1\right)
\tilde d^*_R\tilde d_L+h.c.}\eqn\intb$$
When the neutral Higgs bosons develop  VEVs, the interactions of
eq.~\inta\ and eq.~\intb\ induce complex mass terms for the gauginos,
higgsinos, quarks and squarks.\foot{Neutral Higgs complex mass
 terms are also
induced. For small $\varphi$, however,
they do not give rise to any significant
phenomenological implication.}
Since the phases $\rho$ and
$\xi$ cannot be rotated away, CP is violated by the  fermion and scalars
propagators.
The gauginos
 and higgsinos mix with each other, and the
resulting mass eigenstates are called charginos ($\tilde \chi^+$)
and neutralinos ($\tilde \chi^0$).
Notice that our
model is similar to the supersymmetric model with explicit
CP violation. However, there are some important differences.
The KM matrix is now real\refmark\bra\ as is the gluino
mass ($m_g$). Moreover, all CP violating phases, $\varphi$,
can be written as a function
of only the two phases $\rho$ and $\xi$, \ie,
 $\varphi=\varphi(\rho,\xi)$.
\bigskip
\noindent{\bf 3. Contributions to $\bold\varepsilon$,
$\bold\varepsilon^\prime$ and the NEDM}

\REF\co{L.L. Chau, Phys. Rep. {\bf 95} (1983) 1.}
\REF\expb{J.M. G\'erard,
a talk given in the 15th Int. Symp. on Lepton-Photon
Interactions at High Energies (Geneva, 1991).}
In this section we calculate the predictions of the model described
in the previous section
for the $\varepsilon$ and $\varepsilon^\prime$ parameters and the
NEDM.
 Following the notation
 of ref.~[\co], we have
 $$\varepsilon=\coeff{1}{\sqrt {2}}e^{i\pi/4}\left(\half
t_m+t_0\right)\, ,\eqn\alexf$$
$$\varepsilon^\prime=-\coeff{1}{\sqrt{2}}ie^{i(\delta_2-\delta_0)}\coeff
{{\rm Re}A_2}{{\rm Re}A_0}t_0\, ,\eqn\alexg$$
where $A_i$ are the weak-decay amplitudes of the neutral kaon to two
pions of isospin $i$, $\delta_i$ are
 the corresponding phases from strong
 interactions and
$$t_i=\coeff{{\rm Im}A_i}{{\rm Re}A_i}\, ,\ \ \ t_m=\coeff{{\rm Im}M_{12}
}{{\rm Re}M_{12}}\, ,\eqn\alexh$$
where $M_{ij}$ is
the neutral kaon mass matrix in the $K^0$--$\bar K^0$ basis.
We have used the phase convention such that $t_2=0$.
{}From the experimental values\refmark{\exp,\expb}
$$\eqalign{&|\varepsilon|\simeq \, 2.26\cdot 10^{-3}\, ,\cr
 &\varepsilon^\prime/\varepsilon\lsim\, 1.45\cdot 10^{-3}\, ,\cr
 &|A_2/A_0|\simeq\,  1/22\, ,\cr
 &\delta_2-\delta_0\simeq -53^0\, ,}\eqn\alexi$$
we have
$$\eqalign{t_m&\simeq\, 2\sqrt{2}|\varepsilon|\simeq 6\cdot 10^{-3}\, ,
\cr
 t_0&\simeq\, \sqrt{2} \left|\coeff{A_0}{A_2}\right|
 |\varepsilon^\prime|\lsim  10^{-4}\, .}\eqn\exp$$

\FIG\diao{Dominant one-loop contribution to Im$M_{12}$. We denote by
$\bullet$  a $\tilde H$--$\tilde W$ or  $\tilde t_L$--$\tilde t_R$
  mixing.}
\FIG\diat{Dominant one-loop contribution to Re$M_{12}$.}
To begin with, let us consider the contribution to $t_m$.
 The only diagrams that  considerably  contribute
 to  Im$M_{12}$ are those
 involving phases in the propagators of the superpartners.
In order to have a complex $\tilde\chi^0$ or $\tilde\chi^+$
propagator, it is easy to see from eq.~\inta\ that
mixing between gauginos and higgsinos is required.
As a result, the superbox diagrams must involve a quark-squark-higgsino
coupling so that their contribution to  Im$M_{12}$  is suppressed
by a factor $m_q/m_W$. If the phases arise
from a squark propagator, then  $\tilde q_L$--$\tilde q_R$ mixing is
necessary [see eq.~\intb].
In this case, superbox diagrams
receive a suppression factor  $m_q/m_{\tilde q}$.
Thus, only
superbox diagrams involving $\tilde t$ are nonnegligible.
The largest of these
contributions arise from the diagrams shown in fig.~\diao.
 In addition, there are two more diagrams
 like those of fig.~\diao\ but with  $\tilde H$ and $\tilde W$
interchanged in the $\tilde H$--$\tilde W$ fermion line,
 and contributing an opposite phase. Since these diagrams
involve different super-KM  matrix elements,
 there will be only a partial
cancellation, which we denote by $S$.
 In order that the contribution of the diagrams of fig.~\diao\
be large enough, we need a small mass  for the lightest chargino and
squark. In such a case, the contribution to Re$M_{12}$ given
 by the diagram shown in
fig.~\diat\ is also large and
a degeneracy between $\tilde u_L$  and $\tilde d_L$ is required
 in order to be consistent with the
experimental value.\foot{The contribution to Re$M_{12}$ from
 superbox diagrams involving neutralinos and
gluinos  can be
neglected if the masses of these particles are larger than
$200$ GeV\refmark\fcnct.} From ref.~[\ellis], we have,
for  $m_{\tilde \chi^+}\sim m_{\tilde q}\sim 100$ GeV,
$$\coeff{\Delta m^2_{\tilde q}}{m^2_{\tilde q}}\lsim \calo\left(
\coeff{1}{30}\right)\, .\eqn\bound$$
When the bound \bound\ is saturated, the value of $t_m$ is given by the
ratio between the diagrams of
fig.~\diao\ and fig.~\diat. A rough calculation gives
$$t_m\simeq\coeff{\coeff{m_t}{\sqrt{2}m_W\sin\beta}
V_{13}S\sin\varphi}{\Delta m^2_{\tilde q}/m^2_{\tilde q}}
\, ,\eqn\tm$$
where $\tan\beta=v_2/v_1$ and $V_{13}$ is, according to eq.~\alexb,
 of $\calo (10^{-2})$. We have assumed the maximal
 $\tilde H$--$\tilde  W$ and $\tilde t_L$--$\tilde t_R$ mixing.
This is a natural assumption for $m_{\tilde \chi^+}\sim m_t
\sim 100$ GeV.
The fact that the diagrams of
fig.~\diao\ involve the $\bar dP_R\tilde H^c\tilde t_R$ coupling
which is proportional to $\coeff{m_t}{\sqrt{2}m_W\sin\beta}$
is crucial: the super-GIM mechanism does not apply and such diagrams
receive only one power of the suppression factor $\Delta m^2_{\tilde q}
/m^2_{\tilde q}$.
 For $\tan\beta\simeq 1$, $m_t\simeq 2m_W$ and
$S\simeq 1/2$, we have
$$t_m\simeq 3\cdot 10^{-1}\sin\varphi\, .\eqn\alexj$$
Since this is the maximal contribution to $t_m$, we have from eq.~\exp\
the lower bound
$$\varphi\gsim 2\cdot 10^{-2}\, .\eqn\lb$$

\REF\penguin{M.A. Shifman, A.I. Vainshtein and V.I. Zakharov,
Nucl. Phys. {\bf B120} (1977) 316;\hfill\break
F.J. Gilman and M.B. Wise, Phys. Lett. {\bf B83} (1979) 83.}
\FIG\diath{Dominant one-loop contribution to Im$A_0$. We denote by
$\bullet$  a $\tilde H$--$\tilde W$ or
 $\tilde t_L$--$\tilde t_R$  mixing.}
\FIG\diaf{Chargino one-loop contribution to Re$A_0$.}
\REF\lan{P. Langacker and B. Sathiapalan, Phys. Lett. {\bf B144} (1984)
395.}
\REF\buras{G. Buchalla, A.J. Buras and M.K. Harlander, Nucl. Phys.
{\bf B337} (1990) 313.}
To estimate $t_0$, we assume that $A_0$ is dominated by the penguin
diagrams\refmark\penguin.
The largest contribution to Im$A_0$
arises from penguin diagrams involving charginos and top squarks
 (fig.~\diath).
The chargino penguin diagrams also contribute to Re$A_0$. The dominant
contribution is shown in fig.~\diaf\ and
leads to the effective lagrangian
(for $m_{\tilde\chi^+}\sim m_{\tilde q}$)\refmark\lan
$$\call_{sp} \simeq\coeff{\alpha_s\alpha_W}{24m^2_{\tilde q}}
\sin\theta_c\coeff{\Delta m^2_{\tilde q}}
{m^2_{\tilde q}}{\bf O}_{LR}+h.c.\, ,\eqn\alexk$$
where ${\bf O}_{LR}=(\bar s_L\gamma_\mu T^ad_L)(\bar q_R\gamma^\mu
T^aq_R)$, and $T^a$ is the hermitian SU(3)$_c$ generator.
Considering only the chargino contribution, the ratio $t_0$
 is found to be
 of the same order of the ratio $t_m$ given in eq.~\tm. However,
the dominant  contribution to Re$A_0$ arises  from
the standard penguin  diagram:
$$\call_{p} \simeq\coeff{\alpha_s\alpha_W}{3m^2_W}
\sin\theta_c\ln\coeff{m^2_c}{m^2_K}{\bf O}_{LR}+h.c.\eqn\alexl$$
Therefore,
$$t_0\simeq\coeff{m^2_W}{16m^2_{\tilde q}}
\coeff{\coeff{m_t}{\sqrt{2}m_W\sin\beta}
V_{13}S\sin\varphi}{\ln (m^2_c/m^2_K)}\, .\eqn\alexm$$
For the same values of the parameters considered in obtaining
eq.~\alexj,
we have $t_0\sim 6\cdot 10^{-6}$
in agreement with the experimental limit given in eq.~\exp.
We must remark that the predictions for $t_0$ have large
uncertainties\refmark\buras\
and  cannot be considered a precision test of the model.

\REF\nedmexp{I.S. Altarev \etal, JETP Lett. {\bf 44} (1986) 460;
\hfill\break
K.F. Smith \etal, Phys. Lett. {\bf B234} (1990) 191.}
\FIG\diafi{One-loop chargino diagram
contributing to the NEDM. We denote by a cross a mass insertion in the
fermion line.}
Constraints from the NEDM, $d_n$,
 are more severe.  The predictions of our model for $d_n$
 can be estimated using
previous  calculations of the NEDM
in  supersymmetric models with explicit CP violation.
Such calculations can be found in refs.~[\nedma--\nedmc];
more recent analyses
are given in refs.~[\nedmd--\nedmf].
The dominant contribution  to $d_n$ arises from
  diagrams involving gluinos.
Although the CP violating phase that appears in such diagrams
 is different from the phase that appears in fig.~\diao, both are
of the same order (assuming no accidental cancellation).
    Different contributions
to the NEDM arising from the induced
quark electric dipole moment,  quark
chromo-electric dipole moment, Weinberg's three-gluon operator
and  one-photon-three-gluon
operator have been considered in ref.~[\nedme].
Using the experimental value\refmark\nedmexp\
 $|d_n|<1.2\cdot 10^{-25}$ e cm,
the tightest bound found in ref.~[\nedme] for $m_g\sim m_{\tilde q}\sim
m_Z$  is $\varphi\lsim  7.5\cdot 10^{-3}$.

The next most
important contribution to the NEDM comes from diagrams involving
charginos (e.g., fig.~\diafi). For $m_{\tilde d}\sim m_{\tilde \chi^+}$,
this contribution is given by\refmark\nedmb
 $$d_{n}\simeq \coeff{eg^2}
 {36\sqrt{2}\pi^2m_W\cos\beta}\, \coeff{m_d}{m_{\tilde d}}
 \sin\varphi\eqn\alexn$$
which also gives
 rise to an upper bound for $\varphi$ of $\calo (10^{-2})$.
The fact that these bounds are so close to that of eq.~\lb\   suggests
 that a more  rigorous
 calculation should be carried out in order to determine whether this
 model is ruled
 out. Notice, however, that we still have enough freedom in the
parameter space to decrease the contribution to the NEDM
 without decreasing the contribution  to $t_m$ coming from the diagrams
  of fig.~\diao.
 For example, constraints on $\varphi$ from
 gluino contributions to the NEDM
can be made less severe by taking a larger gluino mass (if
 $m_g\gsim 300$ GeV,
the bound relaxes to $\varphi\lsim 10^{-1}$).
In the chargino case, contributions to the NEDM are smaller
in the region of small $\tan\beta$ or
small soft-supersymmetry-breaking
gaugino mass term, $M< m_W$.\foot{Note that in the limit
$M\rightarrow 0$ the phases in  diagram shown in
\diafi\ can be rotated away
giving a zero contribution to the NEDM.}
In addition, contributions to $t_m$
depend strongly on the super-KM  matrices [see eq.~\tm]
which are in principle arbitrary.
\bigskip
\noindent{\bf 4. Conclusions}

Even if $\bar\theta$ is small due to a PQ symmetry, a massless quark or
some other possible
mechanism,  supersymmetric theories must still face the problem of
having additional CP
violating phases that induce a too large $d_n$.

In this paper we have proposed a supersymmetric model where CP is broken
spontaneously. For this purpose, an additional scalar
 singlet is required. Phases in the VEVs of the neutral Higgs bosons
are then responsible for all CP violating phenomena. They induce
at low energy  complex mass matrices for squarks, charginos and
neutralinos.
The main contribution to $\varepsilon$ arises from superbox diagrams
(fig.~\diao). We showed that if $\Delta m^2_{\tilde q}/m^2_{\tilde q}$
 saturates the
bound derived from experimental limits
from FCNC processes, such diagrams can explain the
experimental value of $\varepsilon$. Contributions to
 $\varepsilon^\prime/\varepsilon$ and the NEDM  are in agreement with
the experimental bounds if the gluino mass is larger than about
$200$ GeV and
the CP violating phases are of $\calo ( 10^{-2})$. The smallness
of these phases is natural in the sense of 't Hooft.

\REF\bb{I.I. Bigi and F. Gabbiani, Nucl. Phys. {\bf B352} (1991) 309.}
Deviations from the SM predictions are expected to be important in
CP violating  B decays. Such  processes will be crucial for
 revealing the detailed structure of this
 model. An analysis of the impact of different classes of
supersymmetric models on B decays can be found in ref.~[\bb].

\REF\kra{L.M. Krauss and S.J. Rey, Preprint NSF-ITP-92-03 (1992).}
We must admit that our model suffers from the usual domain wall
 problems just like most models with SCPV.
A possible solution which avoids such problems
has been recently suggested in ref.~[\kra].

Finally, it is interesting to note that
the above analysis shows
that CP violation
can arise generically
as a supersymmetric effect at low energy. In other words, the
 KM phase is
 not strictly necessary to explain the experimental  observed
 CP violating phenomena.
However, such a picture is consistent only
for a small region of the parameter space of the
supersymmetric model.

\vskip .5cm
\centerline{\bf Acknowledgements}

I would like to thank F. del Aguila, M. Dine, H. Haber and Y. Nir
for helpful discussion and comments.
I am also grateful to M. Dine and H. Haber
for a critical reading of the manuscript.
This work was supported by a fellowship of the MEC (Spain).
\endpage
\refout
\endpage
\figout
\end